\documentclass
[aps,12pt,a4paper,amsfonts,amssymb,titlepage,showpacs,preprint,showkeys,nofootinbib,nobibnotes,byrevtex,superscriptaddress]{revtex4}%
\usepackage{amsfonts}
\usepackage{amsmath}
\usepackage{amssymb}
\usepackage{graphicx}%
\providecommand{\U}[1]{\protect\rule{.1in}{.1in}}
\begin{document}
\title{Gauge Anomaly Cancellation in Chiral Gauge Theories}
\author{Gabriel Di Lemos Santiago Lima}
\email{gabriellemos3@hotmail.com}
\affiliation{Departamento de F\'{\i}sica, Pontif\'{\i}cia Universidade Cat\'{o}lica do Rio
de Janeiro, Rua Marqu\^{e}s de S\~{a}o Vicente 225, Rio de Janeiro, RJ,
22451-900, Brazil}
\author{Rafael Chaves}
\email{rafael.chaves@icfo.es}
\affiliation{Institut de Ci\`{e}ncies Fot\`{o}niques, Av. Carl Friedrich Gauss 3,
Castelldefels, Barcelona, 08860, Spain }
\author{Sebasti\~{a}o Alves Dias}
\email{tiao@cbpf.br}
\altaffiliation{corresponding author}

\affiliation{Centro Brasileiro de Pesquisas F\'{\i}sicas, Rua Dr. Xavier Sigaud 150, Rio de
Janeiro, RJ, 22290-180, Brazil}

\begin{abstract}
We consider chiral fermions interacting minimally with abelian and non-abelian
gauge fields. Using a path integral approach and exploring the consequences of
a mechanism of symmetry restoration, we show that the gauge anomaly has null
expectation value in the vacuum for both cases (abelian and non-abelian). We
argue that the same mechanism has no possibility to cancel the chiral anomaly,
what eliminates competition between chiral and gauge symmetry at full quantum
level. We also show that the insertion of the gauge anomaly in arbitrary gauge
invariant correlators gives a null result, which points towards anomaly
cancellation in the subspace of physical state vectors.

\end{abstract}
\maketitle

\section{Introduction}

A gauge anomaly is the quantum breakdown of gauge invariance
\cite{jackiw,gaume-ginsparg,gaume-witten}. It manifests itself through a
non-null expectation value of the divergence of the gauge current. It appears
in a great variety of contexts, from superstrings \cite{supercordas}, passing
through quantum gravity \cite{robinson} to the description of \ the fractional
quantum Hall effect \cite{tiao-gabriel}. In this work we will consider the
important example of gauge theories of Weyl fermions minimally coupled to
gauge fields. In this situation, the appearance of a gauge anomaly is viewed
as unavoidable due to the necessarily simultaneous occurrence of chiral and
gauge symmetry at classical level and their quantum competition
\cite{weinberg,zumino,stora,bonora}. It is usually said that the gauge anomaly
destroys Slavnov-Taylor identities, crucial for renormalization, and turns
unitarity uncertain. This is enough to discard theories where gauge anomalies
appear, when it is not possible to cancel them through any other means
\cite{modelo-padrao}.

In the discussions mentioned above, it must be noticed that while the anomaly
is a \textit{quantum} phenomenon, it is usually computed as a functional of
the gauge fields, which are then considered as \textit{classical}. This means,
in a path integral context, that one does not usually integrate over them.
However, during the 80's, some works considered the full quantum nature of the
theory (integrating also over the gauge fields) and gave support to the idea
that anomalous gauge theories are not necessarily inconsistent. The pioneering
work was that of Jackiw and Rajaraman \cite{JR} in which it was shown that a
two-dimensional gauge anomalous theory was unitary and had massive photons in
its spectrum. It was soon followed by the one of Faddeev and Shatashvilli
\cite{FS} who noticed that the gauge anomaly could be dealt with by the
introduction of new quantum degrees of freedom, that transformed second class
constraints (correlated to the gauge anomaly) into first class ones.
Integrating over these extra fields and over the chiral fermions, one was led
to an effective action, a functional of the gauge field, which was a
\textit{gauge invariant} one. Then, it was understood independently by Harada
and Tsutsui \cite{HT} and Babelon, Schaposnik and Viallet \cite{BSV} that the
application of Faddeev-Popov's method to an anomalous theory introduced these
new degrees of freedom naturally, associated to the non-factorization of the
integration over the gauge group. These last arguments were no longer
restricted to two dimensions.

In the context of abelian theories in two dimensions, one could see more
recently \cite{casana-tiao} that important issues such as renormalizability
and unitarity could be achieved in gauge anomalous models, both in the vector
(Dirac fermions) and in the chiral case (Weyl fermions), when the gauge field
is also quantized. The fact that the anomaly was a trivial (in the vector
case) or a non-trivial cocycle (in the chiral case) did not seem to matter for
the consistency of the model.

It would be natural to consider what happens to the gauge anomaly in this new
context, of gauge invariant effective actions, in an arbitrary number of
dimensions $d$. One would expect, on the basis of the results obtained in the
80's, that the gauge anomaly should vanish after considering the gauge field
as a \textit{quantum} field. Following this line of reasoning, in this work we
briefly review the approach mentioned above to the gauge anomaly through the
use of functional methods, that incorporate in a natural way the extra degrees
of freedom. Then, using this formalism, we show the vanishing of the vacuum
expectation value (v.e.v.) of the gauge anomaly and of its insertions in
arbitrary gauge invariant correlators for chiral gauge theories in $d$ dimensions.

We organize the discussion as follows: in the second section, we review the
arguments contained in \cite{HT} to obtain gauge invariant effective actions
for anomalous gauge theories, fixing our conventions and definitions in the
process. This is a review section, intended to recall the methods and
procedures used to show symmetry restoration. The third section is devoted to
the consideration of the abelian case in an arbitrary number of dimensions: we
derive an expression for the gauge transform of the gauge anomaly and we show
that the v.e.v. of the gauge anomaly has to vanish as a consequence of it. In
the fourth section, we show that the same argument can not be used for the
non-abelian case, again in $d$ dimensions. By employing a different line of
reasoning we consider the covariant divergence of the gauge current (in terms
of the matter fields) in the fifth section and we show that its v.e.v. has to
vanish. Consistence of our results is checked by indicating how to show
independently that the expression of the v.e.v. of the gauge anomaly in terms
of the effective action (that is, as a functional of the gauge field) also
vanishes. Arbitrary gauge invariant correlators with insertions of the gauge
anomaly are shown to be zero in the sixth section. We discuss the fate of the
chiral anomaly in the seventh section, where we indicate that it remains
possibly different from zero. We present our conclusions in the eighth
section. A small appendix is dedicated to reviewing the proof that the v.e.v.
of an abelian gauge field vanishes.

\section{Quantum restoration of gauge symmetry}

In this section, we briefly review the appearance of a gauge anomaly and we
show the way to restore gauge invariance on an anomalous chiral gauge theory,
along the lines of the work of Harada and Tsutsui \cite{HT}. This will fix our
definitions and conventions, which will be used along the body of our work.

We consider theories described by an action $I[\psi,\bar{\psi},A_{\mu}]$,
given by
\begin{equation}
I[\psi,\bar{\psi},A_{\mu}]=I_{G}[A_{\mu}]+I_{F}[\psi,\bar{\psi},A_{\mu}]=\int
dx\,\frac{1}{2}\text{tr\thinspace}F_{\mu\nu}F^{\mu\nu}+\int dx\,\bar{\psi
}D\psi, \label{action}%
\end{equation}
where $dx$ indicates integration over a $d$-dimensional Minkowski space. The
fields $\psi$ are Weyl fermions carrying the fundamental representation of
$SU(N)$. As usual, $A_{\mu}$ takes values in the Lie algebra of $SU(N)$ such
that%
\begin{equation}
A_{\mu}=A_{\mu}^{a}T_{a},\qquad F_{\mu\nu}=\partial_{\mu}A_{\nu}-\partial
_{\nu}A_{\mu}+ie\left[  A_{\mu},A_{\nu}\right]  ,
\end{equation}
and the generators $T_{a}$ satisfy
\begin{equation}
\left[  T_{a},T_{b}\right]  =if_{abc}T_{c},\qquad\text{tr\thinspace}\left(
T_{a}T_{b}\right)  =-\frac{1}{2}\delta_{ab}.
\end{equation}
The operator $D$ is the covariant derivative, and is called the Dirac operator
of the theory. It is given by
\begin{equation}
D=i\gamma^{\mu}\left(  \partial_{\mu}\mathbf{1}+ieA_{\mu}\right)  \equiv
i\gamma^{\mu}D_{\mu}.
\end{equation}
Under gauge transformations,
\begin{equation}
g=\exp(i\theta^{a}\left(  x\right)  T_{a}),
\end{equation}
and simultaneous changes of the fields $\psi$ and $A_{\mu}$ as%
\begin{align}
A_{\mu}^{g}  &  =gA_{\mu}g^{-1}+\frac{i}{e}\left(  \partial_{\mu}g\right)
g^{-1},\nonumber\\
\psi^{g}  &  =g\psi,\nonumber\\
\bar{\psi}^{g}  &  =\bar{\psi}g^{-1},\,
\end{align}
the action $I$ is classically gauge invariant%
\begin{equation}
I[\psi^{g},\bar{\psi}^{g},A_{\mu}^{g}]=I[\psi,\bar{\psi},A_{\mu}].
\end{equation}
Invariance of the action under gauge transformations leads to the classical
covariant conservation of the \textit{gauge current}%
\begin{equation}
\left(  D_{\mu}\right)  _{ab}J_{b}^{\mu}=0, \label{cons-curr}%
\end{equation}
with
\begin{equation}
J_{a}^{\mu}\equiv\bar{\psi}\gamma^{\mu}T_{a}\psi
\end{equation}
and%
\begin{equation}
\left(  D_{\mu}\right)  _{ab}=\delta_{ab}\partial_{\mu}+ef_{abc}A_{\mu}^{c}.
\end{equation}
The quantum theory is defined by the generating functional, which is
\begin{equation}
Z[\eta,\overline{\eta},j_{a}^{\mu}]=\int d\psi d\bar{\psi}dA_{\mu}\exp\left(
iI[\psi,\bar{\psi},A_{\mu}]+i\int dx[\overline{\eta}\psi+\bar{\psi}\eta
+j_{a}^{\mu}A_{\mu}^{a}]\right)  . \label{gen-func}%
\end{equation}
Non-invariance of the fermion measure under gauge transformations leads to a
potential quantum violation of the classical conservation law (\ref{cons-curr}%
). To see this, we perform the following infinitesimal change of variables
\begin{align}
\psi^{g}  &  =g\psi\approx(1+i\delta\theta^{a}T_{a})\psi,\nonumber\\
\bar{\psi}^{g}  &  =\bar{\psi}g^{-1}\approx\bar{\psi}(1-i\delta\theta^{a}%
T_{a}).
\end{align}
In this way
\begin{align}
Z_{g}  &  =\int d\psi^{g}d\bar{\psi}^{g}dA_{\mu}\exp\left(  iI[\psi^{g}%
,\bar{\psi}^{g},A_{\mu}]+i\int dx\left[  \overline{\eta}\psi^{g}+\bar{\psi
}^{g}\eta+j_{a}^{\mu}A_{\mu}^{a}\right]  \right) \nonumber\\
&  =\int d\psi d\bar{\psi}dA_{\mu}J[A_{\mu},g]\exp\left(  iI[\psi,\bar{\psi
},A_{\mu}]+i\int dx\left[  \overline{\eta}\psi+\bar{\psi}\eta+j_{a}^{\mu
}A_{\mu}^{a}\right]  \right. \nonumber\\
&  \left.  +i\int dx\left(  i\delta\theta^{a}\left[  i\left(  D_{\mu}\right)
_{ab}J_{b}^{\mu}+\overline{\eta}T_{a}\psi-\bar{\psi}T_{a}\eta\right]  \right)
\right)  .
\end{align}
We notice the appearance of a Jacobian $J[A_{\mu},g]=J\left[  A_{\mu}%
,\delta\theta\right]  \equiv\exp(i\alpha_{1}[A_{\mu},\delta\theta])$. Given
the infinitesimal character of the transformation, it can be functionally
expanded to first order in $\delta\theta$
\begin{equation}
J[A_{\mu},\delta\theta]=1+\int dx\,\delta\theta^{a}\mathcal{A}_{a}(A_{\mu
})+...,
\end{equation}
or, in terms of $\alpha_{1}$%
\[
\alpha_{1}[A_{\mu},\delta\theta]=-i\int dx\,\delta\theta^{a}\mathcal{A}%
_{a}(A_{\mu})+....
\]
Imposing that the result of the integral should be the same for both
variables, we have
\begin{equation}
Z=Z_{g},
\end{equation}
which means that
\begin{align}
&  \int d\psi d\bar{\psi}dA_{\mu}dx\exp\left(  iI[\psi,\bar{\psi},A_{\mu
}]+i\int dx\left[  \overline{\eta}\psi+\eta\bar{\psi}+j_{a}^{\mu}A_{\mu}%
^{a}\right]  \right) \nonumber\\
&  \times\,i\delta\theta^{a}[i\left(  D_{\mu}\right)  _{ab}J_{b}^{\mu
}-i\mathcal{A}_{a}(A_{\mu})+\overline{\eta}T_{a}\psi-\bar{\psi}T_{a}%
\eta]\nonumber\\
&  =0.
\end{align}
Then, setting the external sources to zero, we see that
\begin{align}
&  \int d\psi d\bar{\psi}dA_{\mu}\left\{  \left(  D_{\mu}\right)  _{ab}%
J_{b}^{\mu}\right\}  \exp(iI[\psi,\bar{\psi},A_{\mu}])\nonumber\\
&  =\int d\psi d\bar{\psi}dA_{\mu}\left\{  \mathcal{A}_{a}(A_{\mu})\right\}
\exp(iI[\psi,\bar{\psi},A_{\mu}]),
\end{align}
or, in terms of v.e.v.s,
\begin{equation}
\left\langle 0\left\vert \left(  D_{\mu}\right)  _{ab}J_{b}^{\mu}\right\vert
0\right\rangle =\left\langle 0\left\vert \mathcal{A}_{a}(A_{\mu})\right\vert
0\right\rangle . \label{anomaly equation}%
\end{equation}
So, from the functional integral point of view, it has long been clear that a
possible anomaly in gauge symmetry is intrinsically related to the
non-invariance of the fermionic measure \cite{fujikawa}. However, we notice
that \textit{there is still an expectation value to be taken, before we
definitely say that current conservation is violated}.

Before computing $\left\langle 0\left\vert \mathcal{A}_{a}(A_{\mu})\right\vert
0\right\rangle $ it is instructive to look closely at the symmetry structure
of the theory in the absence of gauge invariance of the fermionic measure. It
is well known \cite{faddeev-popov}, in the context of a non-anomalous theory,
that integration over the field $A_{\mu}$ has to be restricted to
configurations that are not physically equivalent, due to the gauge symmetry
of the action. The Faddeev-Popov technique exposes the factorization of the
gauge group volume and restricts integration over non-equivalent
representatives of gauge orbits. Coming back to (\ref{gen-func}) we notice
that, if we proceed applying Faddeev-Popov's method, the gauge volume does not
factor out, since there is an additional dependence on the group elements
coming from the Jacobian,
\begin{equation}
d\psi d{\bar{\psi}}=\exp(i\alpha_{1}[A_{\mu},g^{-1}])d\psi^{g}d\bar{\psi}^{g}.
\end{equation}
Introducing the famous \textquotedblleft$1$\textquotedblright\ of
Faddeev-Popov,%
\begin{equation}
1=\Delta_{\text{FP}}\left[  A_{\mu}\right]  \int dg\,\delta\left(  f\left[
A_{\mu}^{g}\right]  \right)
\end{equation}
in the vacuum amplitude $Z\left[  0\right]  $ (with $\Delta_{\text{FP}}\left[
A_{\mu}\right]  $ being the Faddeev-Popov determinant and $f\left(  A_{\mu
}\right)  =0$ being the gauge fixing condition) we see that%
\begin{align}
Z[0]  &  =\int d\psi d\bar{\psi}dA_{\mu}dg\,\Delta_{\text{FP}}\left[  A_{\mu
}\right]  \delta\left(  f\left[  A_{\mu}^{g}\right]  \right)  \exp\left(
iI\left[  \psi,\bar{\psi},A_{\mu}\right]  \right) \nonumber\\
&  =\int d\psi d\bar{\psi}dA_{\mu}^{g^{-1}}dg\,\Delta_{\text{FP}}\left[
A_{\mu}^{g^{-1}}\right]  \delta\left(  f\left[  \left(  A_{\mu}^{g^{-1}%
}\right)  ^{g}\right]  \right)  \exp\left(  iI\left[  \psi,\bar{\psi},A_{\mu
}^{g^{-1}}\right]  \right) \nonumber\\
&  =\int d\psi d\bar{\psi}dA_{\mu}dg\,\Delta_{\text{FP}}\left[  A_{\mu
}\right]  \delta\left(  f\left[  A_{\mu}\right]  \right)  \exp\left(
iI\left[  \psi^{g},\bar{\psi}^{g},A_{\mu}\right]  \right) \nonumber\\
&  =\int d\psi d\bar{\psi}dA_{\mu}dg\,\Delta_{\text{FP}}\left[  A_{\mu
}\right]  \delta\left(  f\left[  A_{\mu}\right]  \right)  \exp\left(
iI\left[  \psi,\bar{\psi},A_{\mu}\right]  +i\alpha_{1}\left[  A_{\mu}%
,g^{-1}\right]  \right) \nonumber\\
&  =\int d\psi d\bar{\psi}\mathcal{D}A_{\mu}dg\,\exp\left(  iI\left[
\psi,\bar{\psi},A_{\mu}\right]  +i\alpha_{1}\left[  A_{\mu},g^{-1}\right]
\right)  ,
\end{align}
where we defined%
\[
\mathcal{D}A_{\mu}=dA_{\mu}\Delta_{\text{FP}}\left[  A_{\mu}\right]
\delta\left(  f\left[  A_{\mu}\right]  \right)  .
\]
The non-factorization of the gauge volume naturally generates new degrees of
freedom, the \textit{Wess-Zumino fields} $\theta^{a}\left(  x\right)  $, that
come from the integration over the local group element $g\left(  x\right)  $
\begin{equation}
dg=\prod\limits_{a}d\theta_{a}(x).
\end{equation}
Following the spirit of Ref. \cite{HT}, we show below that these degrees of
freedom produce a new gauge invariant action. To see this, we define
\begin{equation}
\exp(iW[A_{\mu}]):=\int d\psi d\bar{\psi}\exp\left(  iI\left[  \psi,\bar{\psi
},A_{\mu}\right]  \right)  .
\end{equation}
The Jacobian can be related to $W\left[  A_{\mu}\right]  $ in the following
way:
\begin{align}
\exp(iW[A_{\mu}^{g}])  &  =\int d\psi d\bar{\psi}\exp\left(  i\int
dx\,\bar{\psi}D(A_{\mu}^{g})\psi\right) \nonumber\\
&  =\int d\psi d\bar{\psi}\exp\left(  i\int dx\,\bar{\psi}^{g^{-1}}D(A_{\mu
})\psi^{g^{-1}}\right) \nonumber\\
&  =J\left[  A_{\mu},g\right]  \exp(iW[A_{\mu}])\nonumber\\
&  \rightarrow J\left[  A_{\mu},g\right]  =\exp\left(  i\left[  W[A_{\mu}%
^{g}]-W[A_{\mu}]\right]  \right)  .
\end{align}
So, we see that $\alpha_{1}$ is given by
\begin{equation}
\alpha_{1}[A_{\mu},g]=W[A_{\mu}^{g}]-W[A_{\mu}],
\end{equation}
and exhibits clearly its behavior under gauge transformations (cocycle
property) \cite{FS2}
\begin{align}
\alpha_{1}[A_{\mu}^{h},g]  &  =W[A_{\mu}^{hg}]-W[A_{\mu}^{h}]\nonumber\\
&  =\alpha_{1}[A_{\mu},hg]-\alpha_{1}[A_{\mu},h].
\end{align}

In particular, we find a familiar expression \cite{gaume-witten} for the
anomaly in terms of $W[A_{\mu}]$:%
\begin{align}
\mathcal{A}_{a}\left(  x\right)   &  =i\left.  \frac{\delta\alpha_{1}\left[
A_{\mu},g\right]  }{\delta\theta_{a}\left(  x\right)  }\right\vert _{\theta
=0}=i\left.  \frac{\delta W\left[  A_{\mu}^{g}\right]  }{\delta\theta
_{a}\left(  x\right)  }\right\vert _{\theta=0}\nonumber\\
&  =\left.  i\int dz\frac{\delta W\left[  A_{\mu}^{g}\right]  }{\delta
A_{\mu,b}^{g}\left(  z\right)  }\frac{\delta A_{\mu.b}^{g}\left(  z\right)
}{\delta\theta_{a}\left(  x\right)  }\right\vert _{\theta=0}\nonumber\\
&  =i\int dz\frac{\delta W\left[  A_{\mu}\right]  }{\delta A_{\mu,b}\left(
z\right)  }\left(  \left.  \frac{\delta A_{\mu,b}^{g}\left(  z\right)
}{\delta\theta_{a}\left(  x\right)  }\right\vert _{\theta=0}\right)  \text{.}%
\end{align}
Using that%
\begin{equation}
\left.  \frac{\delta A_{\mu,b}^{g}\left(  z\right)  }{\delta\theta_{a}\left(
x\right)  }\right\vert _{\theta=0}=-\frac{1}{e}\left(  D_{\mu}\right)
_{ab}\delta\left(  z-x\right)  \label{28}%
\end{equation}
we obtain\footnote{This expression of the gauge anomaly in terms of the
effective action means that we are considering the \textit{consistent}
anomaly, as defined, for example, in section 14.2 of \cite{abdalla-rothe}.}%
\begin{equation}
\mathcal{A}_{a}\left(  x\right)  =\frac{i}{e}\left(  D_{\mu}\frac{\delta
W\left[  A_{\mu}\right]  }{\delta A_{\mu}\left(  z\right)  }\right)  _{a}.
\end{equation}

Now we define an effective action integrating only over the fermions and
Wess-Zumino fields:%
\begin{align}
\exp(iI_{\text{eff}}[A_{\mu}]) &  :=\int d\psi d\bar{\psi}dg\exp(iI[\psi
,\bar{\psi},A_{\mu}]+i\alpha_{1}[A_{\mu},g^{-1}])\nonumber\\
&  =\int d\psi d\bar{\psi}dg\exp(iI[\psi,\bar{\psi},A_{\mu}]+iW[A_{\mu
}^{g^{-1}}]-iW[A_{\mu}])\nonumber\\
&  =\int dg\exp(iW[A_{\mu}^{g^{-1}}]).\label{def-effective}%
\end{align}
The original vacuum amplitude is written in terms of it as%
\begin{equation}
Z\left[  0\right]  =\int\mathcal{D}A_{\mu}\exp\left(  iI_{\text{eff}}[A_{\mu
}]\right)  .\label{tiao-effective}%
\end{equation}
This new effective action is gauge invariant, as is shown below:%
\begin{align}
\exp(iI_{\text{eff}}[A_{\mu}^{h}]) &  =\int dg\exp(iW[\left(  A_{\mu}%
^{h}\right)  ^{g^{-1}}])=\int dg\exp(iW[A_{\mu}^{\left(  h^{-1}g\right)
^{-1}}])\nonumber\\
&  =\int d\left(  h^{-1}g\right)  \exp(iW[A_{\mu}^{\left(  h^{-1}g\right)
^{-1}}])=\exp(iI_{\text{eff}}[A_{\mu}]).
\end{align}
Expression (\ref{tiao-effective}) is the usual one that corresponds to a gauge
theory in which one chooses a gauge fixing condition. Gauge invariance of the
effective action strongly indicates the cancellation of the anomaly, as long
as gauge symmetry is restored at quantum level, through the introduction of
the Wess-Zumino fields. We will verify that this is possibly true in the next sections.

\section{Weyl fermions interacting with an abelian gauge field in $d$
dimensions}

Now we will focus in the v.e.v. of the anomaly. We restrict ourselves to the
abelian case (gauge group $U\left(  1\right)  $, only one generator $T=1$, one
parameter $\theta\left(  x\right)  $, $g=\exp\left(  i\theta\left(  x\right)
\right)  $, $f_{abc}=0$) in an arbitrary number of dimensions in this section.
The v.e.v. of the anomaly can be written as%
\begin{align}
\left\langle 0\left\vert \mathcal{A}(A_{\mu})\right\vert 0\right\rangle  &
=\int d\psi d\bar{\psi}dA_{\mu}\left(  {\mathcal{A}}(A_{\mu})\right)
\exp(iI\left[  \psi,\bar{\psi},A_{\mu}\right]  )\nonumber\\
&  =\int d\psi d\bar{\psi}dA_{\mu}d\theta{\Delta}_{\mathrm{FP}}[A_{\mu}%
]\delta(f(A_{\mu}^{g}))\left(  {\mathcal{A}}(A_{\mu})\right)  \exp\left(
iI\left[  \psi,\bar{\psi},A_{\mu}\right]  \right) \nonumber\\
&  =\int d\psi d\bar{\psi}\mathcal{D}A_{\mu}d\theta\left(  {\mathcal{A}%
}(A_{\mu}^{g^{-1}})\right)  \exp(iI\left[  \psi,\bar{\psi},A_{\mu}\right]
+i\alpha_{1}\left[  A_{\mu},g^{-1}\right]  ),
\end{align}
where we performed the usual steps of the Faddeev-Popov method but took into
consideration that the fermionic measure is not invariant under a gauge
transformation. We notice that the gauge transform of the anomaly can be
written as a functional derivative%
\begin{align}
\mathcal{A}(A_{\mu}^{g^{-1}})  &  =\frac{i}{e}\partial_{\mu}\frac{\delta
W\left[  A_{\mu}^{g^{-1}}\right]  }{\delta A_{\mu}^{g^{-1}}\left(  x\right)
}=-\frac{i}{e}\int dz\frac{\delta W\left[  A_{\mu}^{g^{-1}}\right]  }{\delta
A_{\mu}^{g^{-1}}\left(  z\right)  }\partial_{\mu}\delta\left(  z-x\right)
\nonumber\\
&  =i\int dz\frac{\delta W\left[  A_{\mu}^{g^{-1}}\right]  }{\delta A_{\mu
}^{g^{-1}}\left(  z\right)  }\frac{\delta A_{\mu}^{g^{-1}}\left(  z\right)
}{\delta\theta\left(  x\right)  }=i\frac{\delta W\left[  A_{\mu}^{g^{-1}%
}\right]  }{\delta\theta\left(  x\right)  }=i\frac{\delta\alpha_{1}\left[
A_{\mu},g^{-1}\right]  }{\delta\theta\left(  x\right)  }.
\end{align}
Now we can proceed, using the result just derived
\begin{align}
\left\langle 0\left\vert \mathcal{A}(A_{\mu})\right\vert 0\right\rangle  &
=i\int d\psi d\bar{\psi}\mathcal{D}A_{\mu}d\theta\left(  i\frac{\delta}%
{\delta\theta\left(  x\right)  }\alpha_{1}\left[  A_{\mu},g^{-1}\right]
\right) \nonumber\\
&  \times\exp(iI\left[  \psi,\bar{\psi},A_{\mu}\right]  +i\alpha_{1}\left[
A_{\mu},g^{-1}\right]  )\nonumber\\
&  =\int d\psi d\bar{\psi}\mathcal{D}A_{\mu}d\theta\frac{\delta}{\delta
\theta\left(  x\right)  }[\exp(iI\left[  \psi,\bar{\psi},A_{\mu}\right]
+i\alpha_{1}\left[  A_{\mu},g^{-1}\right]  )]\nonumber\\
&  =\int d\theta\frac{\delta}{\delta\theta\left(  x\right)  }F\left[
\theta\right]  =0.
\end{align}
which shows that the anomaly vanishes because of the translational invariance
of the functional measure \cite{rivers}.

We would like to briefly comment on the special case $d=2$, where the gauge
anomaly is \cite{JR}%
\begin{equation}
\mathcal{A}\left(  A_{\mu}\right)  =-\frac{e}{4\pi}\left\{  \left(
a-1\right)  \partial_{\mu}A^{\mu}+\varepsilon^{\mu\nu}\partial_{\mu}A_{\nu
}\right\}  .
\end{equation}
It is a linear function of the gauge field and so it is obvious that its
v.e.v. vanishes, as a consequence of Poincar\'{e} invariance of the vacuum:%
\begin{equation}
\left\langle 0\left\vert A_{\mu}\left(  x\right)  \right\vert 0\right\rangle
=0. \label{vevamu}%
\end{equation}
In fact, again considering the $2$-dimensional case, this is true also for the
non-abelian case, because the anomaly is given by
\begin{equation}
\mathcal{A}_{a}\left(  A_{\mu}\right)  =-\frac{e}{4\pi}\left\{  \left(
a-1\right)  \partial_{\mu}A_{a}^{\mu}+\varepsilon^{\mu\nu}\partial_{\mu
}A_{a,\nu}\right\}  .
\end{equation}
For completeness of the argument, we will briefly present a demonstration of
equation (\ref{vevamu}) in an appendix.

\section{The non-abelian case}

Inspired by the mechanism of anomaly cancellation in the abelian case, we
investigate if it is possible to generalize it for the non-abelian situation
\cite{teserafael}. Again, the focus is the v.e.v. of the anomaly, which is
given, as before, by the expression%
\begin{align}
\left\langle 0\left\vert \mathcal{A}(A_{\mu})\right\vert 0\right\rangle  &
=\int d\psi d\bar{\psi}dA_{\mu}\left(  {\mathcal{A}}(A_{\mu})\right)
\exp(iI\left[  \psi,\bar{\psi},A_{\mu}\right]  )\nonumber\\
&  =\int d\psi d\bar{\psi}dA_{\mu}d\theta{\Delta}_{\mathrm{FP}}[A_{\mu}%
]\delta(f(A_{\mu}))\left(  {\mathcal{A}}(A_{\mu}^{g^{-1}})\right)
\exp(iI\left[  \psi,\bar{\psi},A_{\mu}\right]  +i\alpha_{1}\left[  A_{\mu
},g^{-1}\right]  ).
\end{align}
Starting again from the expression%
\begin{equation}
\mathcal{A}_{a}\left(  x\right)  =\frac{i}{e}\left(  D_{\mu}\frac{\delta
W\left[  A_{\mu}\right]  }{\delta A_{\mu}\left(  z\right)  }\right)  _{a}.
\end{equation}
it is easy to write%
\begin{equation}
\mathcal{A}_{a}(A_{\mu}^{g^{-1}})=\frac{i}{e}\left(  D_{\mu}^{g^{-1}}%
\frac{\delta W\left[  A_{\mu}^{g^{-1}}\right]  }{\delta A_{\mu}^{g^{-1}%
}\left(  x\right)  }\right)  _{a}=-\frac{i}{e}\int dz\frac{\delta W\left[
A_{\mu}^{g^{-1}}\right]  }{\delta A_{\mu,b}^{g^{-1}}\left(  z\right)  }\left(
D_{\mu}^{g^{-1}}\right)  _{ba}\delta\left(  z-x\right)  .
\end{equation}
In order to proceed along the same lines, we have to investigate if the
following equation is true or not%
\begin{equation}
-\frac{i}{e}\left(  D_{\mu}^{g^{-1}}\right)  _{ba}\delta\left(  z-x\right)
=\lambda\frac{\delta A_{\mu,b}^{g^{-1}}\left(  z\right)  }{\delta\theta
^{a}\left(  x\right)  },\label{analytic}%
\end{equation}
with $\lambda$ being a constant (independent of $\theta^{a}$, to be
determined). If it is true, then the v.e.v. of the non-abelian anomaly cancels
with an argument which is parallel to that of the abelian case. This is not an
easy question to be answered, in the general case ($SU\left(  N\right)  $). We
will analyse the case in which the gauge group is $SU\left(  2\right)  $. In
the fundamental representation, the generators $T_{a}$ are given by%
\begin{equation}
T_{1}=\frac{1}{2}\left(
\begin{array}
[c]{cc}%
0 & 1\\
1 & 0
\end{array}
\right)  \text{,\qquad}T_{2}=\frac{1}{2}\left(
\begin{array}
[c]{cc}%
0 & -i\\
i & 0
\end{array}
\right)  \text{,\qquad}T_{3}=\frac{1}{2}\left(
\begin{array}
[c]{cc}%
1 & 0\\
0 & -1
\end{array}
\right)  \text{.}%
\end{equation}
\ The gauge group element has the well known closed form%
\begin{equation}
g\left(  \theta\right)  =\exp\left(  i\theta^{a}T_{a}\right)  =\cos\left(
\frac{\theta}{2}\right)  +2in^{a}T_{a}\sin\left(  \frac{\theta}{2}\right)
\text{,}\label{analytic1}%
\end{equation}
with%
\begin{align}
\theta &  =\sqrt{\left(  \theta^{1}\right)  ^{2}+\left(  \theta^{2}\right)
^{2}+\left(  \theta^{3}\right)  ^{2}},\\
n^{a} &  =\frac{\theta^{a}}{\theta}.\nonumber
\end{align}
\textit{Equation (\ref{analytic1}) is valid only for} $\theta\neq0$. This is
an important observation and we will return to it in the end of the
calculation. Some auxiliary results will be useful during this computation:%
\begin{align*}
\partial_{\mu}\theta &  =n^{a}\partial_{\mu}\theta^{a},\\
\partial_{\mu}n^{a} &  =\frac{1}{\theta}\left(  \delta^{ab}-n^{a}n^{b}\right)
\partial_{\mu}\theta^{b},\\
\left(  n^{a}T_{a}\right)  \left(  \partial_{\mu}n^{b}T_{b}\right)   &
=\frac{1}{2\theta}\left(  \vec{n}\times\partial_{\mu}\vec{\theta}\right)
^{c}T_{c},\\
\frac{\delta\theta\left(  z\right)  }{\delta\theta^{a}\left(  x\right)  } &
=n^{a}\left(  z\right)  \delta\left(  z-x\right)  ,\\
\frac{\delta n^{b}\left(  z\right)  }{\delta\theta^{a}\left(  x\right)  } &
=\frac{1}{\theta}\left(  \delta^{ba}-n^{b}n^{a}\right)  \left(  z\right)
\delta\left(  z-x\right)  ,
\end{align*}
The gauge transform of the gauge field can be put into a completely analytical
form. It is not difficult to obtain the two contributions for the gauge
transform of the gauge field:%
\begin{equation}
A_{\mu}^{g^{-1}}=g^{-1}A_{\mu}g+\frac{i}{e}\left(  \partial_{\mu}%
g^{-1}\right)  g,\label{ag1}%
\end{equation}%
\begin{equation}
g^{-1}A_{\mu}g=A_{\mu}+\,\left(  \vec{n}\times\vec{A}_{\mu}\right)  ^{a}%
\sin\theta T_{a}-2\,\left(  \left(  \vec{n}\times\vec{A}_{\mu}\right)
\times\vec{n}\right)  ^{a}\sin^{2}\left(  \frac{\theta}{2}\right)
T_{a},\label{ag2}%
\end{equation}%
\begin{equation}
\frac{i}{e}\left(  \partial_{\mu}g^{-1}\right)  g=\frac{1}{e}n^{a}\left(
\partial_{\mu}\theta\right)  T_{a}-\frac{1}{e\theta}\left(  \partial_{\mu}%
\vec{n}\times\vec{n}\right)  ^{a}+\frac{1}{e}\left(  \partial_{\mu}%
n^{a}\right)  \sin\,\theta T_{a}+\frac{1}{e\theta}\left(  \partial_{\mu}%
\vec{n}\times\vec{n}\right)  ^{a}\,\cos\theta T_{a}.\label{ag3}%
\end{equation}
To proceed, we should functionally differentiate the expression above with
respect to $\theta^{a}\left(  x\right)  $ and compare the result with $\left(
D_{\mu}^{g^{-1}}\right)  _{ba}\delta\left(  z-x\right)  $. This is a long and
tedious task. We can circumvent this calculation if we perform a consistency
check. By computing%
\[
\frac{\delta A_{\mu,b}^{g^{-1}}\left(  z\right)  }{\delta\theta^{a}\left(
x\right)  }=\frac{\delta}{\delta\theta^{a}\left(  x\right)  }\left[  \left(
g^{-1}A_{\mu}^{c}T_{c}g\right)  ^{b}+\frac{i}{e}\left(  \left(  \partial_{\mu
}g^{-1}\right)  g\right)  ^{b}\right]
\]
one should be able to obtain, among other terms, something proportional to the
first term of the covariant derivative%
\[
\delta^{ba}\partial_{\mu}\delta\left(  z-x\right)  .
\]
So, we can investigate the possible contributions to a term like this in the
functional derivative. The term $g^{-1}A_{\mu}g$ clearly can not contribute as
it contains no derivatives of $\theta^{a}$. The last two terms in the
expression of $\left(  \partial_{\mu}g^{-1}\right)  g$ will always contain
$\sin\theta$ and $\cos\theta$ and so they are not candidates. All that remains
is to compute the contributions of the first two terms:%
\begin{align*}
\frac{\delta}{\delta\theta^{a}\left(  x\right)  }\left(  \frac{1}{e}%
n^{b}\left(  \partial_{\mu}\theta\right)  \left(  z\right)  \right)   &
=\frac{1}{e}n^{b}\left(  z\right)  \left(  \partial_{\mu}\frac{\delta
\theta\left(  z\right)  }{\delta\theta^{a}\left(  x\right)  }\right)  +...\\
&  =\frac{1}{e}\left(  n^{a}n^{b}\right)  \left(  z\right)  \partial_{\mu
}\delta\left(  z-x\right)  +...,
\end{align*}%
\begin{align*}
\frac{\delta}{\delta\theta^{a}\left(  x\right)  }\left(  -\frac{\varepsilon
_{bcd}}{e\theta}\left(  \partial_{\mu}n^{c}\right)  n^{d}\left(  z\right)
\right)   &  =-\frac{\varepsilon_{bcd}}{e\theta}\left(  \partial_{\mu}%
\frac{\delta n^{c}\left(  z\right)  }{\delta\theta^{a}\left(  x\right)
}\right)  n^{d}\left(  z\right)  +...\\
&  =-\frac{\varepsilon_{bcd}}{e\theta}\partial_{\mu}\left(  \frac{1}{\theta
}\left(  \delta^{ca}-n^{c}n^{a}\right)  \left(  z\right)  \delta\left(
z-x\right)  \right)  n^{d}\left(  z\right)  +...\\
&  =-\frac{\varepsilon_{bcd}}{e\theta^{2}}\left(  n^{d}\delta^{ca}-n^{c}%
n^{a}n^{d}\right)  \left(  z\right)  \partial_{\mu}\delta\left(  z-x\right)
+...
\end{align*}
This reasoning proves that $\delta^{ba}\partial_{\mu}\delta\left(  z-x\right)
$ is not contained in the expression for $\delta A_{\mu,b}^{g^{-1}}\left(
z\right)  /\delta\theta^{a}\left(  x\right)  $, so it can not be proportional
to $\left(  D_{\mu}^{g^{-1}}\right)  _{ba}\delta\left(  z-x\right)  $. This
means that the v.e.v. of the gauge anomaly, in the non-abelian case, does not
cancel in the same way as in the abelian case. We stress that we did not
proved that the v.e.v. of the gauge anomaly is non-zero for the non-abelian
case. We proved that it is not cancelled by a generalization of the argument
given in section 3. However, in the next sections, we will give more general
arguments to show that it cancels also in the non-abelian case.

Before ending this section we want to compare our findings with equation
(\ref{28}). It is precisely the relation we were seeking (changing $g$ for
$g^{-1}$), but computed for $\theta^{a}=0$. A generalization of this equation
would be natural (and would be equation (\ref{analytic})) and then one could
think to have arrived at a contradiction. In fact, the exact analytic form for
$A_{\mu}^{g^{-1}}$ (expressions (\ref{ag1}), (\ref{ag2}) and (\ref{ag3})) is
valid only under the assumption that some of the $\theta^{a}$ are distinct
from zero. So, (\ref{28}) would not be obtainable from (\ref{analytic}) (if it
would be true) and, in this way, we could suspect that the generalization of
(\ref{28}) for $\theta^{a}\neq0$ would not be trivial. The exact analytic form
of $\delta A_{\mu,b}^{g^{-1}}\left(  z\right)  /\delta\theta^{a}\left(
x\right)  $ can be directly (but patiently) obtained from the elements we gave
and we see that the expression is far more complicated than the (relatively)
simple form $\left(  D_{\mu}^{g^{-1}}\right)  _{ba}\delta\left(  z-x\right)
$. What makes the non-abelian case so different from the abelian one remains
to be discovered.

\section{Null divergence of the gauge current}

Now we will give general arguments to the cancellation of v.e.v. of the gauge
anomaly even in the non-abelian case, in $d$ dimensions. We will consider the
v.e.v. of the covariant divergence of the current and we will show very simply
that it has to vanish. To see this, we consider a bosonic change of variables
in the functional integral:%
\begin{align}
Z[0]  &  =\int d\psi d\bar{\psi}dA_{\mu}\exp\left(  iI[\psi,\bar{\psi},A_{\mu
}]\right)  =\int d\psi d\bar{\psi}dA_{\mu}^{g}\exp\left(  iI[\psi,\bar{\psi
},A_{\mu}^{g}]\right) \nonumber\\
&  =\int d\psi d\bar{\psi}dA_{\mu}\exp\left(  iI[\psi,\bar{\psi},A_{\mu}%
^{g}]\right)  , \label{translacao}%
\end{align}
where we used again the invariance of the bosonic measure $dA_{\mu}%
^{g}=dA_{\mu}$. The functional integral does not contain a gauge group volume,
as it would happen in a non-anomalous gauge theory, because fermion
integration produces a gauge non-invariant $W\left[  A_{\mu}\right]  $. So,
the Faddeev-Popov trick is unnecessary here, as it has already been stressed.
This way of facing the problem is known as \textit{gauge non-invariant
representation} \cite{abdalla-rothe,cesar-rothe}.

Next, we consider an infinitesimal gauge transformation, characterized by
$g\approx1+i\delta\theta^{a}T_{a}$,%
\begin{align}
I[\psi,\bar{\psi},A_{\mu}^{g}]  &  =I[\psi,\bar{\psi},A_{\mu}+\frac{1}%
{e}D_{\mu}\delta\theta]\nonumber\\
&  =I[\psi,\bar{\psi},A_{\mu}]+\frac{1}{e}\int dx\,\left(  D_{\mu}\delta
\theta\left(  x\right)  \right)  _{a}\frac{\delta I}{\delta A_{\mu}^{a}\left(
x\right)  }\nonumber\\
&  =I[\psi,\bar{\psi},A_{\mu}]-\frac{1}{e}\int dx\,\delta\theta^{a}\left(
x\right)  \left(  D_{\mu}\frac{\delta I}{\delta A_{\mu}\left(  x\right)
}\right)  _{a}.
\end{align}
This gives%
\begin{align}
Z[0]  &  =\int d\psi d\bar{\psi}dA_{\mu}\exp\left(  iI[\psi,\bar{\psi},A_{\mu
}^{g}]\right) \nonumber\\
&  \approx\int d\psi d\bar{\psi}dA_{\mu}\exp\left(  iI[\psi,\bar{\psi},A_{\mu
}]\right) \nonumber\\
&  -\frac{1}{e}\int dx\,\delta\theta^{a}\left(  x\right)  \int d\psi
d\bar{\psi}dA_{\mu}\left(  D_{\mu}\frac{\delta I}{\delta A_{\mu}\left(
x\right)  }\right)  _{a}\exp\left(  iI[\psi,\bar{\psi},A_{\mu}]\right)
\nonumber\\
&  \Rightarrow\int d\psi d\bar{\psi}dA_{\mu}\left(  D_{\mu}\frac{\delta
I}{\delta A_{\mu}\left(  x\right)  }\right)  _{a}\exp\left(  iI[\psi,\bar
{\psi},A_{\mu}]\right)  =0.
\end{align}
Remembering that%
\begin{equation}
\frac{\delta I}{\delta A_{\nu}^{a}\left(  x\right)  }=\left(  D_{\mu}F^{\mu
\nu}\right)  _{a}-e\bar{\psi}\gamma^{\nu}T_{a}\psi,
\end{equation}
and that $\left(  D_{\mu}D_{\nu}F^{\mu\nu}\right)  _{a}=0$ identically,%
\begin{equation}
\left(  D_{\mu}\frac{\delta I}{\delta A_{\mu}\left(  x\right)  }\right)
_{a}=\left(  D_{\mu}\right)  _{ab}\bar{\psi}\gamma^{\mu}T_{b}\psi,
\end{equation}
we conclude that%
\begin{equation}
\left\langle 0\left\vert \left(  D_{\mu}\right)  _{ab}\bar{\psi}\gamma^{\mu
}T_{b}\psi\right\vert 0\right\rangle =\left\langle 0\left\vert \left(  D_{\mu
}\right)  _{ab}J_{b}^{\mu}\right\vert 0\right\rangle =0.
\end{equation}
We notice that this result was reached without making any fermionic change of
variables. In fact, it does not even matter if the fermionic measure is gauge
invariant or not. The above equation is also completely consistent with our
previous conclusions in the abelian case, but it is also valid for the
non-abelian case as well.

We can face this result as a general argument (i.e. independent of the
consideration of Weyl or Dirac fermions) for the vanishing of the expectation
value of the covariant divergence of the gauge current. It relies only on the
existence of classical gauge symmetry and gauge invariance of the gauge field
functional measure. It means that, if the theory is to be consistent, the
right-hand side of equation (\ref{anomaly equation}) \textit{must vanish
accordingly}. In the case where we are considering Dirac fermions, this is
seen as a trivial consequence of the invariance of the fermion measure under
gauge transformations (which says that the Jacobian $J[A_{\mu},g]=1$ and thus
that $\mathcal{A}_{a}=0$). However, in the Weyl fermion case, the Jacobian is
not trivial and we must investigate in further detail if the right hand side
vanishes or if we simply ended at an inconsistency. This can be easily
achieved in two steps: first, gauge invariance of $I_{\text{eff}}$ (equation
(\ref{def-effective})) implies%
\[
\int dg\left(  \delta\theta^{g^{-1}}\right)  ^{a}(x)\left[  D_{\mu}^{g^{-1}%
}\left(  \frac{\delta W[A_{\mu}^{g^{-1}}]}{\delta A_{\mu}^{g^{-1}}(x)}\right)
\right]  _{a}\exp\left(  iW[A_{\mu}^{g^{-1}}]\right)  =0,
\]
with $\delta\theta^{g^{-1}}=g^{-1}\delta\theta g$ and $\delta\theta
=\delta\theta^{a}T_{a}$. Integrating this equation over the gauge field and
using again the gauge invariance of the bosonic measure we obtain%
\begin{align}
0  &  =\int dA_{\mu}\left[  D_{\mu}\left(  \frac{\delta W[A_{\mu}]}{\delta
A_{\mu}(x)}\right)  \right]  _{a}\exp\left(  iW[A]\right) \\
&  \Rightarrow\left\langle 0\left\vert \mathcal{A}_{a}(A_{\mu})\right\vert
0\right\rangle =0.\nonumber
\end{align}
This works as a consistency check of the anomaly equation, as this vanishing
is obtained on a independent basis of the vanishing of the divergence of the
fermion current (although both proofs use gauge invariance of $dA_{\mu}$).

\section{The gauge anomaly and the Hilbert space}

A very important question is if the behaviour found above for the anomaly is
repeated for its insertion on an arbitrary correlation function. If this would
be true it would be a very dangerous situation for the theory, for it could
imply that the gauge field operator has to be zero itself, what could mean
that the theory is trivial (or inconsistent). We found no means to answer
completely this question. However, it can be answered in the case of arbitrary
gauge invariant correlators, as we show below.

Any gauge invariant operators can be expressed in terms of $\psi$, $\bar{\psi
}$ and $A_{\mu}$. The basic property that they must obey is%
\begin{equation}
O\left(  \psi^{g},\bar{\psi}^{g},A_{\mu}^{g}\right)  =O\left(  \psi,\bar{\psi
},A_{\mu}\right)  .
\end{equation}
This property, by its turn, implies that%
\begin{equation}
O\left(  \psi,\bar{\psi},A_{\mu}^{g}\right)  =O\left(  \psi^{g^{-1}},\bar
{\psi}^{g^{-1}},A_{\mu}\right)  .
\end{equation}
That is all we need to analyse gauge invariant correlators. We consider the
generating functional for correlators of general composite operators
$O_{i}\left(  \psi,\bar{\psi},A_{\mu}\right)  $,%
\begin{equation}
Z_{O}\left[  \lambda_{i}\right]  =\int d\psi d\bar{\psi}dA_{\mu}\exp\left(
iI[\psi,\bar{\psi},A_{\mu}]+i\int dx\lambda_{i}O_{i}\left(  \psi,\bar{\psi
},A_{\mu}\right)  \right)  ,
\end{equation}
Arbitrary correlators are obtained from $Z_{O}\left[  \lambda_{i}\right]  $ in
a standard way:%
\begin{equation}
\left\langle 0\right\vert T\left(  O_{i_{1}}\left(  x_{1}\right)  ...O_{i_{n}%
}\left(  x_{n}\right)  \right)  \left\vert 0\right\rangle =\left.  \left(
-i\right)  ^{n}\frac{\delta^{n}Z_{O}\left[  \lambda_{i}\right]  }%
{\delta\lambda_{i_{1}}\left(  x_{1}\right)  ...\delta\lambda_{i_{n}}\left(
x_{n}\right)  }\right\vert _{\lambda=0}.
\end{equation}
Now, we can begin integrating over $A_{\mu}^{g}$ instead of $A_{\mu}$ (as
before, a dummy integration variable):
\begin{align}
Z_{O}\left[  \lambda_{i}\right]   &  =\int d\psi d\bar{\psi}dA_{\mu}^{g}%
\exp\left(  iI[\psi,\bar{\psi},A_{\mu}^{g}]+i\int dx\lambda_{i}O_{i}\left(
\psi,\bar{\psi},A_{\mu}^{g}\right)  \right) \nonumber\\
&  =\int d\psi d\bar{\psi}dA_{\mu}\exp\left(  iI[\psi^{g^{-1}},\bar{\psi
}^{g^{-1}},A_{\mu}]+i\int dx\lambda_{i}O_{i}\left(  \psi^{g^{-1}},\bar{\psi
}^{g^{-1}},A_{\mu}\right)  \right) \nonumber\\
&  =\int d\psi d\bar{\psi}dA_{\mu}\exp\left(  iI[\psi,\bar{\psi},A_{\mu
}]+i\int dx\lambda_{i}O_{i}\left(  \psi,\bar{\psi},A_{\mu}\right)
-i\alpha_{1}\left(  A_{\mu},g^{-1}\right)  \right)  ,
\end{align}
where we took into account the gauge invariance of the gauge field measure and
the gauge non-invariance of the fermion measure%
\begin{align}
dA_{\mu}  &  =dA_{\mu}^{g},\nonumber\\
d\psi d{\bar{\psi}}  &  =\exp(-i\alpha_{1}[A_{\mu},g^{-1}])d\psi^{g^{-1}}%
d\bar{\psi}^{g^{-1}}.
\end{align}
Restricting ourselves to an infinitesimal gauge transformation,
\begin{equation}
\alpha_{1}\left(  A_{\mu},-\delta\theta\right)  =i\int dx\,\delta\theta
^{a}\mathcal{A}_{a}(A_{\mu})+....,
\end{equation}
we obtain%
\begin{align}
Z_{O}\left[  \lambda_{i}\right]   &  =Z_{O}\left[  \lambda_{i}\right]  -i\int
dx\,\delta\theta^{a}\int d\psi d\bar{\psi}dA_{\mu}\mathcal{A}_{a}(A_{\mu}%
)\exp\left(  iI[\psi,\bar{\psi},A_{\mu}]+i\int dx\lambda_{i}O_{i}\left(
\psi,\bar{\psi},A_{\mu}\right)  \right) \nonumber\\
&  \Rightarrow\int d\psi d\bar{\psi}dA_{\mu}\mathcal{A}_{a}(A_{\mu}%
)\exp\left(  iI[\psi,\bar{\psi},A_{\mu}]+i\int dx\lambda_{i}O_{i}\left(
\psi,\bar{\psi},A_{\mu}\right)  \right)  =0.
\end{align}
Taking arbitrary functional derivatives with respect to $\lambda_{i}$ and
setting them to zero we obtain%
\begin{equation}
\left\langle 0\right\vert T\left(  \mathcal{A}_{a}(A_{\mu})O_{i_{1}}\left(
x_{1}\right)  ...O_{i_{n}}\left(  x_{n}\right)  \right)  \left\vert
0\right\rangle =0.
\end{equation}

We showed that the gauge anomaly has null insertion into arbitrary gauge
invariant correlators. This fact indicates that the anomaly is null in the
subspace of the Hilbert space consisting of physical vectors (those who are
annihilated by the BRST charge, obeying physical restrictions such as gauge
conditions \cite{brst}). It must be remarked that the technique above cannot
say anything about the case of gauge non-invariant correlators. Other
techniques (such as lattice calculations, for example) can bring some light to
this question.

\section{Chiral versus gauge symmetry}

Another important question concerns the chiral anomaly. When the gauge field
is not quantized, it is well known \cite{weinberg,zumino,stora,bonora} that
chiral symmetry is in competition with gauge symmetry. Then, we must
investigate what happens with the chiral anomaly in a context where the gauge
anomaly cancels. Classical chiral symmetry is expressed by the transformations
\cite{modelo-padrao}
\begin{align}
\psi^{c}  &  =c\psi,\nonumber\\
\bar{\psi}^{c}  &  =\bar{\psi}c,\\
A_{\mu}^{c}\,  &  =A_{\mu},
\end{align}
where $c=\exp\left(  i\alpha\gamma_{d+1}\right)  $, with $\partial_{\mu}%
\alpha=0$ and\footnote{The essential property of $\gamma_{d+1}$ is $\left\{
\gamma_{d+1},\gamma^{\mu}\right\}  =0$, which is not satisfied in odd
dimensions. In order to avoid this difficulty, we will restrict our
discussion, in this section, to the case in which $d$ is even.} $\gamma
_{d+1}=i\gamma^{0}\gamma^{1}...\gamma^{d-1}$. The classical action
(\ref{action}) is symmetric with respect to this transformation%
\begin{equation}
I[\psi^{c},\bar{\psi}^{c},A_{\mu}]=I[\psi,\bar{\psi},A_{\mu}],
\end{equation}
and, as a consequence,
\[
\partial_{\mu}J_{d+1}^{\mu}\equiv\partial_{\mu}\left(  \bar{\psi}\gamma^{\mu
}\gamma_{d+1}\psi\right)  =0.
\]
Following the same standard reasoning of section II (an infinitesimal chiral
transformation, with $\partial_{\mu}\delta\alpha\neq0$) we conclude that%
\begin{align}
Z_{c}  &  =\int d\psi^{c}d\bar{\psi}^{c}dA_{\mu}\exp\left(  iI[\psi^{c}%
,\bar{\psi}^{c},A_{\mu}]+i\int dx\left[  \overline{\eta}\psi^{c}+\bar{\psi
}^{c}\eta+j_{a}^{\mu}A_{\mu}^{a}\right]  \right) \nonumber\\
&  =\int d\psi d\bar{\psi}dA_{\mu}J[A_{\mu},c]\exp\left(  iI[\psi,\bar{\psi
},A_{\mu}]+i\int dx\left[  \overline{\eta}\psi+\bar{\psi}\eta+j_{a}^{\mu
}A_{\mu}^{a}\right]  \right. \nonumber\\
&  \left.  +i\int dx\left(  i\delta\alpha\left[  \partial_{\mu}J_{d+1}^{\mu
}+\overline{\eta}\gamma_{d+1}\psi+\bar{\psi}\gamma_{d+1}\eta\right]  \right)
\right)  .
\end{align}
Expanding the Jacobian%
\begin{align}
J[A_{\mu},c]  &  =J[A_{\mu},\delta\alpha]\equiv\exp\left(  i\beta_{1}\left[
A_{\mu},\delta\alpha\right]  \right) \nonumber\\
&  \approx1+i\int dx\,\delta\alpha\left(  x\right)  \left.  \frac{\delta
\beta_{1}}{\delta\alpha\left(  x\right)  }\left(  A_{\mu}\right)  \right\vert
_{\delta\alpha=0}\nonumber\\
&  \equiv1+\int dx\,\delta\alpha\left(  x\right)  \mathcal{A}_{d+1}\left(
A_{\mu}\right)  ,
\end{align}
and imposing $Z=Z_{c}$, we find the usual equation that indicates the possible
presence of the chiral anomaly:%
\begin{equation}
\left\langle 0\left\vert \partial_{\mu}J_{d+1}^{\mu}\right\vert 0\right\rangle
=\left\langle 0\left\vert \mathcal{A}_{d+1}(A_{\mu})\right\vert 0\right\rangle
\label{chiral}%
\end{equation}

We can try to follow the same path as in sections V and VI, in order to see if
we can prove that one or both of the two sides of (\ref{chiral}) is zero. In
section V we began by making a change of variables in the expression for
$Z\left[  0\right]  $ (equation (\ref{translacao})): it was a gauge
transformation on $A_{\mu}$. Then we explored gauge invariance of the bosonic
measure $dA_{\mu}$ to show that the gauge anomaly vanished. As the chiral
transformation affects only the fermion fields (because $A_{\mu}^{c}\,=A_{\mu
}$), it is not possible to apply this procedure here. The same reasoning
invalidates the extension of the procedure followed in the end of section V to
the case of the chiral anomaly, because there is no sense in asking if
$I_{\text{eff}}\left[  A_{\mu}\right]  $ is or is not chiral invariant. So, we
see that one can not conclude that the chiral anomaly vanishes on the basis of
the same arguments that we used to the case of gauge symmetry.

\section{Conclusion}

We considered the anomaly equation%
\[
\left\langle 0\left\vert \left(  D_{\mu}\right)  _{ab}J_{b}^{\mu}\right\vert
0\right\rangle =\left\langle 0\left\vert \mathcal{A}_{a}(A_{\mu})\right\vert
0\right\rangle
\]
and showed arguments to support the vanishing of both sides, independently. In
one case (l.h.s.) we considered the expectation value of a fermionic/bosonic
operator $\left(  D_{\mu}\right)  _{ab}J_{b}^{\mu}$ and showed that it
vanishes. On the other side of the equation, we had the expectation value of a
completely bosonic operator $\mathcal{A}_{a}(A_{\mu})$, which was shown to be
also zero. In both demonstrations we have something in common: gauge
invariance of the \textit{bosonic} measure. So, gauge invariance at quantum
level seems to be determined by the behaviour of the fluctuations of the
$A_{\mu}$ field. The conclusion is that gauge invariance of the fermionic
measure does not seems to be important to guarantee gauge invariance of the
full theory. It seems that gauge invariance is much more resistant than it is
usually supposed.

However, we showed that there are signs indicating that the precise mechanisms
behind the cancelling of the gauge anomaly in the abelian and non-abelian
cases are very distinct. The cancellation of the anomaly in the abelian case
seems to be a consequence of the Dyson-Schwinger equations extended to the
Wess-Zumino fields. This could indicate that the anomaly should be dealt with
as a subsidiary condition (similar to what happens in the case of a Proca
model, when we obtain $\partial_{\mu}A^{\mu}$ as a consistency condition).
This seems not to be the situation in the non-abelian case. What is the true
mechanism behind the cancellation of the non-abelian anomaly is a question
that deserves to be investigated further.

The chiral anomaly was shown not to be cancelled on the basis of the arguments
that we presented. Although the proof is quite trivial, it is very important
to stress this fact explicitly, because this anomaly has deep phenomenological
implications (in an opposite way to the gauge anomaly, which is only used to
find inconsistencies in the theory). It remains possibly different from zero,
both in the vector and in the chiral cases.

The functional formalism points out the origin of the gauge anomaly and a road
for its formal cancellation. Restoration of gauge symmetry implies a null
expectation value for the anomaly. This cancellation suggests that anomalies
are not an obstacle to the quantization of theories involving chiral fermions.
The usual argument is that anomalies destroy Slavnov-Taylor identities,
necessary to relate renormalization constants and prove the renormalizability
of the theory. On the basis of our results, there is no reason to believe that
Slavnov-Taylor identities are not preserved in a gauge anomalous theory. A
detailed analysis of the perturbative renormalization procedure under this new
perspective would be very important and will be considered in detail in the
future. We already began to investigate it by considering insertions of the
gauge anomaly in arbitrary gauge invariant correlators. Their vanishing points
in the direction of a null anomaly in the physical subspace. However, we must
also obtain progress in the gauge non-invariant case, if we want to understand
the full dynamics of the gauge field in a so called gauge anomalous theory.

One can see that the main difference between the vector case (coupling to both
chiralities) and the chiral case (coupling to just one of them) is that gauge
invariance can be maintained at all steps of the quantization in the vector
case, even before quantizing the gauge field. In the chiral case, this is not
so. One has to go through the complete quantization of the theory to see gauge
invariance again. But it seems to be present there, in the end.

\section{Appendix}

We present a simple demonstration of the vanishing of the v.e.v. of the gauge
field. First we show that the derivative of a scalar operator has vanishing
v.e.v.: let $U\left(  \Lambda,a\right)  $ denote the operator that represents
a Poincar\'{e} transformation in Hilbert space ($\Lambda$ is the Lorentz
transformation and $a$ is the vector that indicates translation). Then%
\begin{equation}
\left\langle 0\left\vert \partial_{\mu}\Omega\left(  x\right)  \right\vert
0\right\rangle =\left\langle 0\left\vert U^{\dag}\left(  1,x\right)
\partial_{\mu}\Omega\left(  0\right)  U\left(  1,x\right)  \right\vert
0\right\rangle =\left\langle 0\left\vert \partial_{\mu}\Omega\left(  0\right)
\right\vert 0\right\rangle ,
\end{equation}
as expected from Poincar\'{e} invariance of the vacuum. Now, performing a pure
Lorentz transformation,%
\begin{equation}
\left\langle 0\left\vert \partial_{\mu}\Omega\left(  0\right)  \right\vert
0\right\rangle =\left\langle 0\left\vert U^{\dag}\left(  \Lambda,0\right)
\partial_{\mu}\Omega\left(  0\right)  U\left(  \Lambda,0\right)  \right\vert
0\right\rangle =\Lambda_{\mu}{}^{\nu}\left\langle 0\left\vert \partial_{\nu
}\Omega\left(  0\right)  \right\vert 0\right\rangle .
\end{equation}
This can only be true (for general $\Lambda$) if $\left\langle 0\left\vert
\partial_{\mu}\Omega\left(  0\right)  \right\vert 0\right\rangle =0$. Now, we
can consider the gauge field: using a similar argument for translations we
find%
\begin{equation}
\left\langle 0\left\vert A_{\mu}\left(  x\right)  \right\vert 0\right\rangle
=\left\langle 0\left\vert A_{\mu}\left(  0\right)  \right\vert 0\right\rangle
.
\end{equation}
However, under a general Lorentz transformation, this field does not behaves
as a tensor (see \cite{bjorken})
\begin{equation}
U^{\dag}\left(  \Lambda,0\right)  A_{\mu}\left(  x\right)  \left(  0\right)
U\left(  \Lambda,0\right)  =\Lambda_{\mu}{}^{\nu}A_{\nu}\left(  \Lambda
x\right)  +\partial_{\mu}\Omega\left(  \Lambda,x\right)  ,
\end{equation}
with $\Omega\left(  \Lambda,x\right)  $ being a scalar field dependent on the
particular Lorentz transformation under consideration. So,%
\begin{align}
\left\langle 0\left\vert A_{\mu}\left(  0\right)  \right\vert 0\right\rangle
&  =\left\langle 0\left\vert U^{\dag}\left(  \Lambda,0\right)  A_{\mu}\left(
0\right)  U\left(  \Lambda,0\right)  \right\vert 0\right\rangle \nonumber\\
&  =\Lambda_{\mu}{}^{\nu}\left\langle 0\left\vert A_{\nu}\left(  0\right)
\right\vert 0\right\rangle +\left\langle 0\left\vert \partial_{\mu}%
\Omega\left(  \Lambda,0\right)  \right\vert 0\right\rangle \nonumber\\
&  =\Lambda_{\mu}{}^{\nu}\left\langle 0\left\vert A_{\nu}\left(  0\right)
\right\vert 0\right\rangle =0.
\end{align}

\begin{acknowledgments}
G. L. S. Lima was financially supported by CAPES (Brazil) and R. Chaves was
financially supported by FAPERJ (Brazil), during the realization of this work.
We acknowledge very useful discussions with Edgar Corr\^{e}a de Oliveira and
Jos\'{e} Abdalla Helay\"{e}l Neto (both from CBPF).
\end{acknowledgments}

\end{document}